\documentstyle[titlepage,12pt]{article}
\def\thefootnote{\fnsymbol{footnote}}
\def\@makefnmark{\mbox{$^\@thefnmark$}}
\def\thefootnote{\mbox{\noindent$\fnsymbol{footnote}$}}
\long\def\@makefntext#1{\noindent$^{\@thefnmark}$#1}

\def\appendix{%\vskip 1cm\par\setcounter{equation}{0}
\def\theequation{A1.\arabic{equation}}}

\def\openone{\leavevmode\hbox{\small1\kern-3.3pt\normalsize1}}

\newcommand{\preprint}[2]{\large
\begin{center}
\hspace*{8cm} Preprint  \, EFUAZ\,  FT-94-{#1}\\
\hspace*{8cm} hep-th/9409200\\
\hspace*{8cm}  {#2} 1994
\end{center}}

\begin{document}

\title{\vspace*{-35mm} \preprint{09}{September} \vspace*{2cm}
{\Large {\bf ADDENDUM \linebreak
to the papers on the Weinberg Theory}}\thanks
{Submitted to ``Phys. Lett. B".}}

\author{{\bf Valeri V. Dvoeglazov}\thanks{On leave of absence from
{\it Dept. Theor. \& Nucl. Phys., Saratov State University,
Astrakhanskaya str., 83, Saratov\, RUSSIA}}\,
\thanks{Email: valeri@bufa.reduaz.mx,
dvoeglazov@main1.jinr.dubna.su}\\
%{\em
Escuela de F\'{\i}sica, Universidad Aut\'onoma de Zacatecas \\
Antonio Doval\'{\i} Jaime\, s/n, Zacatecas 98000, ZAC., M\'exico}
%}

\begin{abstract}

The Weinberg-Tucker-Hammer equations are shown to substitute
the common-used $j=1$ massless equations. Meantime, the old
equations preserve their significance as a particular case.

Possible consequences are discussed.
\end{abstract}

\date{\empty}

\maketitle

\renewcommand{\thefootnote}{\arabic{footnote}}

\newpage

The attractive Weinberg's $2(2j+1)$ component formalism for a description
of higher spin particles~\cite{Weinberg}  lately got developed
significantly in connection with the recent works of
Dr.~D.V. Ahluwalia {\it et al.}, ref.~\cite{DVA1}-\cite{DVA3}, and with
the papers~\cite{DVO1}-\cite{DVO4}.

The main equation in this formalism,
which had been proposed in ref.~\cite{Weinberg},
is\footnote{See for discussions on the needed
modifications of this formalism refs.~\cite{DVA3}-\cite{DVO3}
and what follows.}
\begin{equation}\label{eq0}
\left [\gamma_{\mu_1 \mu_2 \ldots \mu_{2j}} p_{\mu_1} p_{\mu_2} \ldots
p_{\mu_{2j}}+ m^{2j}\right ] \psi = 0,
\end{equation}
of the ``$2j$" order in the momentum,
$p_{\mu_i}=-i\partial/\partial x_{\mu_i}$, $m$ is a particle mass. The
analogues of the Dirac $\gamma$- matrices are $2(2j+1)\otimes 2(2j+1)$
matrices which also have  ``$2j$" vectorial indices, ref.~\cite{Barut}.

For the moment I have to repeat the  previous results.
The equations (4.19,4.20), or equivalent to them Eqs. (4.21,4.22),
presented in ref.~[1b,p.B888] and in many other publications:
\begin{eqnarray}
{\bf \nabla}\times \left [ {\bf E} -i {\bf B}\right ]
+i (\partial/\partial t) \left [{\bf E} -i{\bf B} \right ] &=&0, \qquad
(4.21) \nonumber\\
{\bf \nabla}\times \left [ {\bf E} +i {\bf B}\right ]
-i (\partial/\partial t) \left [{\bf E} +i{\bf B} \right ] &=&0, \qquad
(4.22) \nonumber
\end{eqnarray}
are found in ref.~\cite{DVA1} to have acausal
solutions.  Apart from the correct dispersion relations $E=\pm p$ one has
a wrong dispersion relation $E=0$. The origin of this fact is the same to
the problem of the ``relativistic cockroach nest" of
Moshinsky and Del Sol, ref.~\cite{DVO4,Moshinsky}.
On the other hand, {\it "the $m\rightarrow 0$
limit of Joos-Weinberg finite-mass wave equations, Eq. (\ref{eq0}),
satisfied by $(j,0)\oplus (0,j)$ covariant spinors, ref.~\cite{DVA2},
are free from all kinematic acausality."}
The same authors (Dr. D. V. Ahluwalia and his
collaborators) proposed the Bargmann-Wightman-Wigner-type quantum field
theory, {\it ``in which bosons and antibosons  have
opposite relative intrinsic parities"}, ref.~\cite{DVA3}.
This {\it Dirac}-like modification of the Weinberg theory
is an excellent example of combining the Lorentz and the
dual transformations.
In ref.~\cite{DVO1} I concern with the connection between
antisymmetric tensor field models, e.~g.~refs.~\cite{AVD1,AVD2}, and
the equations considered by Weinberg (and by Hammer and Tucker~\cite{Hammer}
in a slightly different form). In the case of the choice\footnote{My
earlier attempts to give a potential interpretation of the $\psi$ were not
successful, ref.~\cite{OLD}.}
\begin{equation}\label{psi}
\psi=\pmatrix{{\bf E}+i{\bf B} \cr {\bf E}-i{\bf B}\cr}
\end{equation}
the equivalence
of the Weinberg-Tucker-Hammer approach and the Proca
approach\footnote{I mean the equations for $F_{\mu\nu}$,
the antisymmetric field tensor, e.~g., Eqs. (3) and (4) in~\cite{DVO1}
or Eq. (A4) in~[15b] which corresponds to a parity conjugated wave function.}
has been found. The necessity of a consideration of another
equation (the Weinberg ``double")\footnote{It is  useful to compare the
method applied in the papers~\cite{DVA1,DVO1}
with the way of deriving the Dirac equation by Dr.~P.~A.~M.~Dirac
himself, ref.~\cite{Dirac}.} was point out. In fact it is
the equation for the antisymmetric tensor dual to $F_{\mu\nu}$, which
had also been considered earlier, e.~g.~ref.~\cite{Takahashi}.
In the paper~\cite{DVO2} the Weinberg fields were
shown to describe the particle
with transversal components (i.~e., of spin~$j=1$) as opposed to
conclusions of refs.~\cite{AVD1,AVD2} and of the previous ones.
The origins of contradictions with the Weinberg theorem ($B-A=\lambda$),
which have been met in the old works (of both mine and others),
have been clarified. The causal propagator of the
Weinberg theory has been proposed in
ref.~\cite{DVO3}. Its remarkable feature is a presence of
four terms. This fact is explained in my forthcoming paper~\cite{DVO4}.

For the moment I am going to consider the question,
under what conditions the Weinberg-Tucker-Hammer
$j=1$ equations can be transformed to Eqs. (4.21) and (4.22)
of ref.~[1b]~? By using the
interpretation of $\psi$, Eq. (\ref{psi}), and the explicit form of
the Barut-Muzinich-Williams matrices, ref.~\cite{Barut},
I am able to recast the $j=1$ Tucker-Hammer equation
\begin{equation}
\left [\gamma_{\mu\nu} p_\mu p_\nu +p_\mu p_\mu +2m^2 \right ] \psi = 0,
\end{equation}
which is free of tachyonic solutions,
or the Proca equation, Eq. (3) in ref.~\cite{DVO1}, to the
form\footnote{I restored $c$, the light velocity, at the
terms.}$^,$\footnote{The reader can recover the dual equations
from  Eqs. (10) or (12) of ref.~\cite{DVO1} without any problems.}
\begin{eqnarray}\label{ME0}
m^2 E_i &=& -{1\over c^2} {\partial^2 E_i \over \partial t^2}
+\epsilon_{ijk} {\partial \over \partial x_j}
{\partial B_k \over \partial t} +
{\partial \over \partial x_i} {\partial E_j \over \partial x_j},\\
\label{ME1}
m^2 B_i &=& {1\over c^2} \epsilon_{ijk} {\partial \over
\partial x_j} {\partial E_k \over \partial t} +  {\partial^2 B_i \over
\partial x_j^2} -{\partial \over \partial x_i} {\partial B_j \over
\partial x_j}.
\end{eqnarray}
The D'Alembert equation (the Klein-Gordon equation in the momentum
representation
indeed)
\begin{equation}\label{DAL}
\left ({1\over c^2} \frac{\partial^2}{\partial t^2} -\frac{\partial^2}{\partial
x_j^2}\right ) F_{\mu\nu}
= -m^2 F_{\mu\nu}
\end{equation}
is implied, ref.~\cite{DVO1}.

Restricting ourselves by the consideration of
the $j=1$ massless case we are able
to re-write them to the following form:
\begin{eqnarray}\label{MY1}
{\partial \over \partial t}\, curl\, {\bf B} + grad\, div\, {\bf E}
-{1\over c^2} {\partial^2 {\bf E} \over \partial t^2} &=& 0,\\
\label{MY2}
{\bf \nabla}^2 {\bf B} -grad\, div\, {\bf B} +{1\over c^2} {\partial \over
\partial t}\, curl\, {\bf E} &=& 0.
\end{eqnarray}
Let me consider the first equation (\ref{MY1}). We can satisfy it
provided that
\begin{equation}\label{E1}
\tilde\rho_e = div \, {\bf E} = const_x \qquad {\bf J} = curl \, {\bf B} -
{1\over c^2} {\partial {\bf E} \over \partial t} = const_t .
\end{equation}
However, this is a particular case only.
Let me mention that the equation
\begin{equation}\label{1}
{\partial {\bf J} \over \partial t} = - grad\, \tilde\rho_e
\end{equation}
follows from (\ref{MY1}) provided that ${\bf J}$ and $\tilde\rho_e$
are defined as in Eq. (\ref{E1}).

Now  I need to take relations of vector algebra in mind:
\begin{equation}
curl\, curl\, {\bf X} = grad\, div\, {\bf X} -{\bf \nabla}^2 {\bf X},
\end{equation}
where ${\bf X}$ is an arbitrary vector.
Recasting Eq. (\ref{MY1})
and taking the D'Alembert equation  (\ref{DAL})  in mind
one can achieve in the general case
\begin{equation}\label{4}
{\bf J^\prime} = {\partial {\bf B} \over \partial t} + curl\, {\bf E} =
 grad\,\chi^\prime,
\end{equation}
in order to satisfy the recasted equation (\ref{MY1}):
\begin{equation}
curl \, {\bf J^\prime} = 0 .
\end{equation}

The second equation (\ref{MY2}) yields
\begin{equation}\label{2}
{\bf J} =  curl \, {\bf B} - {1\over c^2}
{\partial {\bf E} \over \partial t} =  grad \, \chi
\end{equation}
(in order to satisfy $curl\,{\bf J} = 0$).
After adding and subtracting
${1\over c^2} \partial^2 {\bf B} / \partial t^2$
one can obtain
\begin{equation}\label{EQ2}
\tilde \rho_m = div \, {\bf B} = const_x
\qquad{\partial {\bf B} \over \partial t} +curl\, {\bf E} = const_t ,
\end{equation}
provided that
\begin{equation}
{\bf \nabla}^2 {\bf B} -{1\over c^2}
{\partial^2 {\bf B} \over \partial t^2} =0
\end{equation}
(i.~e.~ again the D'Alembert equation taken in account).
The set of equations (\ref{EQ2}), with the constants are chosen to be zero,
are {\it ``an identity satisfied by certain space-time derivatives
of $F_{\mu\nu}$\ldots, namely,\footnote{Ref.~\cite{FD1,FD2}.}}
\begin{equation}
{\partial F_{\mu\nu} \over \partial x^\sigma} +
{\partial F_{\nu\sigma} \over \partial x^\mu} +
{\partial F_{\sigma\mu} \over \partial x^\nu} = 0.
\end{equation}

However, it is also a particular case.
Again, the general solution is
\begin{eqnarray}\label{3}
{1\over c^2} \frac{\partial {\bf J^\prime}}{\partial t} = grad \,
\tilde\rho_m .
\end{eqnarray}
We must pay attention at  the universal case.
What are the functions $\chi$?  {}From Eqs. (\ref{1}) and (\ref{2}) one can
conclude
\begin{equation}
\tilde \rho_e =  \frac{\partial \chi}{\partial t} + const,
\end{equation}
and from (\ref{4}) and (\ref{3}),
\begin{equation}
\tilde \rho_m =  {1\over c^2}\frac{\partial\chi^\prime}{\partial t} + const .
\end{equation}
It is useful to compare the definitions $\tilde\rho_e$ and ${\bf J}$ and the
fact of  an appearance of the functions $\chi$ with the 5-potential
formulation of electromagnetic theory~\cite{FD2}, see also
refs.~\cite{Gersten,FD3}.  I suppose, we have to use  the names
{\it ``prana"}, {\it ``ch'i"} or {\it ``ether"}  for the functions $\chi$.

Finally, I would like to note the following. We can obtain
\begin{eqnarray}
div \, {\bf E} &=& 0 ,\\
{1\over c^2} {\partial {\bf E} \over \partial t}
- curl \, {\bf B} &=& 0 ,\\
div\, {\bf B} &=& 0 ,\\
{\partial {\bf B} \over \partial t} + curl\, {\bf E} &=& 0 ,
\end{eqnarray}
{\it which are just the Maxwell's free-space equations,}
in the definite choice  of the  $\chi$ and $\chi^\prime$, namely,
in the case when they are constants. In the general case of the Weinberg
equations or the Tucker-Hammer equations (the latter possess
the correct dispersion relations
as opposed to the Maxwell equations) the electric charge is conserved
provided that the $\chi$ is a linear function (decreasing?) of the time.

Next, if  I use electric and magnetic fields
as field functions, of course, the question arises on the transformations
from one to other frame (in fact, the question is:
if there is  an unique frame for determinations of  the electric and magnetic
fields or not). The answer can be found easily
in a consideration of the problem in the momentum representation.
For the first sight, one could conclude that
the Lorentz transformations
change the values of $\vec {\cal E}$ and $\vec {\cal B}$ under the transfer
from one to another frame. However, that's wrong!
The invariance follows from the possibility of combining the
Lorentz, dual (chiral) and parity transformations in the case of
higher spin equations\footnote{The equations for the four functions
$\psi_i^{(k)}$, Eqs. (8), (10), (18) and (19) of ref.~\cite{DVO1},
are reduced in the equations for the electric  and  magnetic fields,
which are the same for each case in a massless limit.}.  This possibility
has been discovered and investigated in  refs.~\cite{DVA3,DVO2}.  The four
bispinors $u_1^{\sigma\, (1)} ({\bf p})$, $u_2^{\sigma\, (1)} ({\bf p})$,
$u_1^{\sigma\, (2)} ({\bf p})$ and $u_2^{\sigma\, (2)} ({\bf p})$, see
Eqs. (10), (11), (12) and (13) of ref.~\cite{DVO3}, form the complete set
(as well as $\Lambda ({\bf p}) u_i^{\sigma\, (k)} ({\bf p}))$, namely
\begin{eqnarray}
\lefteqn{c_1 u_1^{\sigma\, (1)} ({\bf p}) \bar u_1^{\sigma\, (1)}
({\bf p}) + c_2
u_2^{\sigma\, (1)} ({\bf p}) \bar u_2^{\sigma\, (1)} ({\bf p})
+\nonumber}\\
&+& c_3 u_1^{\sigma\, (2)} ({\bf p}) \bar u_1^{\sigma\, (2)}
({\bf p}) + c_4 u_2^{\sigma\, (2)} ({\bf p}) \bar u_2^{\sigma\, (2)}
({\bf p}) = \openone .
\end{eqnarray}
The constants $c_i$ are defined by the choice of
the normalization of bispinors.
In any other frame we are able to obtain
the primary wave function by choosing  the appropriate coefficients of the
expansion of the wave function (in fact, by using appropriate dual
rotations and/or inversions)\footnote{The paper which is
devoted to the important experimental consequences of this fact
(e.~g., the Aharonov-Bohm effect  and some {\it others}.) is in progress.} .
Of course, the same conclusion is valid for
negative-energy solutions, since they coincide with the positive-energy
ones in the case of a  $j=1$ boson, ref.~\cite{DVO2,Hammer}.

Finally, let me mention that in the nonrelativistic limit $c\rightarrow
\infty$ one obtains the dual
Levi-Leblond's ``Galilean Electrodynamics",
ref.~\cite{Levy,Crawford}.

For the moment I agree with Dr. S. Weinberg who spoke out about
Maxwell's equations (4.21) and (4.22):
{\it ``The fact that these (!) field
equations are of first order for any spin seems to me to be of no great
significance\ldots"}~[1b, p. B888].

{\it Conclusion}\footnote{This conclusion also follows from the results
of the paper~\cite{DVA1,DVO1,OLD} and ref.~[1b] provided that
the fact that $({\bf j} {\bf p})$ has no inverse one
has been taken into account.}:
The Weinberg-Tucker-Hammer massless equations (or the Proca equations for
$F_{\mu\nu}$), see also (\ref{ME0}) and (\ref{ME1}), are equivalent to the
Maxwell's equations in a definite choice of the initial and boundary
conditions only.  They (Eq. (1) for spin~$j$) was shown in
refs.~\cite{DVA1,DVO1} to be free from all kinematical acausality as opposed to
Eqs. (4.21) and (4.22).  Therefore, in the case of $j=1$ massless
particles (photons) the Weinberg-Tucker-Hammer
equations  substitute the old Maxwell's free-space
equations. Meantime, the old Maxwell's equations preserve their
significance as a particular case. E.~g.,~ in ref.~\cite{Gersten}
it was mentioned: The solutions of Eqs. (4.21,4.22) of ref.~[1b]
satisfy Eqs. of the form (\ref{ME0}) and (\ref{ME1}), {\it ``but not
always vice versa."}

Investigations of Eq. (1) with other initial and boundary conditions
(or of the functions $\chi$) deserve further elaboration (both theoretical
and experimental).

{\it Acknowledgements.} I thank
Profs. A.~Turbiner and Yu.~F.~Smirnov
for the questions on the non-relativistic limit of
the Weinberg-Tucker-Hammer equations, Prof. I. G. Kaplan, for useful remarks,
and all those who gave a initial
impulse to start the work in this direction.

I am grateful to Zacatecas University for a professorship.

This {\it Addendum} has been thought on September 3-4, 1994
%(Shana Tova y metuka!)
as a result of discussions
at the IFUNAM seminar (M\'exico,  2/IX/94).


\begin{thebibliography}{99}

\footnotesize{
\bibitem{Weinberg} S. Weinberg, Phys. Rev.
B133 (1964) B1318; ibid B134 (1964) B882\\[-7mm]
\bibitem{DVA1} D. V. Ahluwalia and D. J. Ernst, Mod. Phys. Lett. A7 (1992)
1967\\[-7mm]
\bibitem{DVA2} D. V. Ahluwalia and D. J. Ernst, Int. J. Mod.
Phys. E2 (1993) 397 \\[-7mm]
\bibitem{DVA3} D. V. Ahluwalia, M. B. Johnson and
T.  Goldman, Phys.  Lett.  B316 (1993) 102\\[-7mm]
\bibitem{DVO1} V. V. Dvoeglazov, {\it Mapping between antisymmetric
tensor and Weinberg formulations.} Preprint EFUAZ FT-94-05
(hep-th/9408077). Zacatecas, M\'exico, August 1994, submitted
to ``Phys. Lett.  B"\\[-7mm]
\bibitem{DVO2} V. V. Dvoeglazov, {\it What particles are described
by the Weinberg theory?} Preprint EFUAZ FT-94-06
(hep-th/9408146). Zacatecas, M\'exico, August 1994, submitted to ``Phys.
Lett. B"\\[-7mm]
\bibitem{DVO3} V. V. Dvoeglazov, {\it The Weinberg propagators.}
Preprint EFUAZ FT-94-07 (hep-th/9408176). Zacatecas,
M\'exico, August 1994, submitted to ``Phys. Lett.
B"\\[-7mm]
\bibitem{DVO4} V. V. Dvoeglazov, {\it Neutrino theory of
light? Part I.} Preprint EFUAZ FT-94-08, Zacatecas, M\'exico, September
1994, submitted to ``Nuovo Cimento B"\\[-7mm]
\bibitem{Barut} A. Barut, I. Muzinich and D. N. Williams,
Phys. Rev. 130 (1963) 442\\[-7mm]
\bibitem{Moshinsky} M. Moshinsky and A. Del Sol, {\it
Relativistic Cockroach Nest.} Preprint IFUNAM FT-93-018, M\'exico,  May
1993, to appear in ``Canadian J. Phys."\\[-7mm]
\bibitem{AVD1} L. V. Avdeev and M. V. Chizhov, Phys.Lett. B321 (1994)
212\\[-7mm]
\bibitem{AVD2} L. V. Avdeev and M. V. Chizhov, {\it A
queer reduction of degrees of freedom.} Preprint JINR
E2-94-263 (hep-th/9407067), Dubna, Russia, July
1994\\[-7mm]
\bibitem{Hammer} R. H. Tucker and C. L. Hammer, Phys. Rev.
D3 (1971) 2448\\[-5mm]
\bibitem{OLD} V. V. Dvoeglazov, Hadronic J.
16 (1993) 423; V. V.Dvoeglazov and S. V. Khudyakov,
Izvest. VUZov:fiz. No. 9 (1994) 110; V. V.
Dvoeglazov, {\it $2(2S+1)$ component model and its connection with
other field theories.} Preprint IFUNAM FT-94-36 (hep-th/9401043),
M\'exico, January 1994, to appear in ``Rev. Mex. Fis."\\[-7mm]
\bibitem{Gersten} A. Gersten, {\it Conserved currents of the Maxwell
equations in the presence of electric and magnetic sources.} Preprint
CERN-TH.4687/87, March 1987; {\it Conserved currents of the Maxwell
equations with electric and magnetic sources.} Preprint CERN-TH.4688/87,
March 1987\\[-7mm]
\bibitem{Dirac} P. A. M. Dirac, Proc. Roy. Soc. A
117 (1928) 610\\[-7mm]
\bibitem{Takahashi} Y. Takahashi and R. Palmer, Phys. Rev.
D1 (1970) 2974\\[-7mm]
\bibitem{FD1} F. J. Dyson, Am. J. Phys. 58
(1990) 209; N. Dombey, ibid 59 (1991) 85; R. W. Brehme, ibid 59
(1991) 85; J. L. Anderson, ibid 59 (1991) 86; I. E. Farquhar, ibid
59 (1991) 87; R. J. Hughes, ibid 60 (1992) 301; C. R. Lee, Phys.
Lett. A 148 (1990) 146; I. E. Farquhar, ibid 151 (1991) 203; A.
Vaidya and C. Farina, ibid 153 (1991) 265 \\[-7mm]
\bibitem{FD2} S. Tanimura, Ann. Phys. (USA) 220 (1992) 229; D. Saad, L. P.
Horwitz and R. I. Arshansky, Found. Phys. 19 (1989) 1126; M. C.
Land, N. Shnerb and L. P.  Horwitz, {\it Feynman's proof of the
Maxwell equations as a way beyond the Standard Model.} Preprint
TAUP-2076-93 (hep-th/9308003), Tel Aviv, Israel\\[-7mm]
\bibitem{FD3} A. Bandyopadhyay, Int. J. Theor. Phys. 32 (1993) 1563\\[-7mm]
\bibitem{Levy} J.-M. Levy-Leblond, Comm. Math. Phys. 6 (1967) 286;
M. Le Bellac and J.-M. Levy-Leblond, Nuovo Cim. B14 (1973) 217\\[-7mm]
\bibitem{Crawford} F. S. Crawford, Am. J. Phys. 60 (1992) 109;
ibid 61 (1993) 472; A. Horzela, E. Kapu\'scik and C. A. Uzes,
ibid 61 (1993) 471\\[-7mm]}

\end{thebibliography}
\end{document}